\documentclass[prl,reprint,aps,superscriptaddress,longbibliography,footinbib]{revtex4-2}

\usepackage{amsmath}
\usepackage{amssymb}
\usepackage[varbb,varg]{newtx} 
\usepackage{microtype}
\usepackage{graphicx}
\usepackage{dcolumn}
\usepackage{bm}
\usepackage{braket}
\usepackage{feynmp}
\usepackage{graphics}
\usepackage{enumerate}
\usepackage[dvipsnames]{xcolor}
\usepackage{comment}
\usepackage[normalem]{ulem} 
\usepackage{hyperref}
\hypersetup{colorlinks=true, linkcolor=NavyBlue, urlcolor=NavyBlue, citecolor=NavyBlue} 
\usepackage[capitalise]{cleveref}

\usepackage{siunitx}

\begin{document}

\title{Charge-ordered states in twisted MoTe$_2$}

\author{Sparsh Mishra}
\email{sparsh@utexas.edu}
\affiliation{Department of Physics, The University of Texas at Austin, Austin, Texas 78712, USA}

\author{Tobias M. R. Wolf}
\affiliation{Department of Physics, The University of Texas at Austin, Austin, Texas 78712, USA}

\author{Allan H. MacDonald}
\affiliation{Department of Physics, The University of Texas at Austin, Austin, Texas 78712, USA}

\date{\today}

\begin{abstract}
We analyze interaction-driven charge-density-wave (CDW) states in the 
spin--valley polarized first valence miniband of twisted MoTe$_2$ (tMoTe$_2$) using an adiabatic mapping from the continuum model to an effective Landau-level (LL) problem. When projected to the lowest LL, the leading spatial harmonic of the
moir\'e-periodic potential changes sign at a magic twist angle $\theta_c$ where the band reaches its minimum bandwidth.  By solving self-consistent Hartree--Fock equations in a 
multi-LL Hilbert space, we find that triangular-lattice CDW states with density maxima
on MX (or XM) sites or on MM sites are favored on opposite sides of the magic angle
at most filling factors and that stripe order appears near $\nu_h=1/2$. 
We show that CDW states at $\nu_h >1/2$ can carry a nonzero total Chern number, providing a natural route to reentrant 
integer quantum Hall effects and discuss the energy competition between 
fractional Chern insulator and CDW states.
\end{abstract}

\maketitle

The recent observations of fractional quantum 
anomalous Hall effects (FQAHEs) in twisted MoTe$_2$ (tMoTe$_2$) homobilayers~\cite{cai2023signatures,park2023observation,xu2023fci,Zeng2023thermodynamic,li2025quantum} 
and in multilayer rhombohedral graphene~\cite{lu2024fractional,nguyen2025hierarchy,lu2025extended} aligned with 
hexagonal boron nitride, have established these systems as platforms for fractional Chern insulator (FCI) states. FCIs emerge 
when strong electron--electron interactions 
dominate the kinetic energy within a topologically nontrivial, nearly flat electronic 
band \cite{neupert2011fractional,sheng2011fractional,wang2011fractional,regnault2011fractional,ledwith2023vortexability,abouelkomsan2020particle,ledwith2020fractional}. Like their magnetic-field-driven counterparts --- fractional quantum Hall effect (FQHE)~\cite{tsui1982two} states --- FCIs compete closely 
with interaction-driven phases that break translational symmetry \cite{reddy2023fractional,li2025multiband}. 
In the FQHE case, electron and hole Wigner crystals~\cite{wigner1934interaction}, bubble phases~\cite{Koulakov_Shklovskii_1996}, and stripe 
phases~\cite{Lilly_Cooper_Eisenstein_Pfeiffer_West_1999} are well-established as 
ground states over some ranges of Landau-level filling~\cite{Yoshioka_Lee_1983,yoshioka1979charge,macdonald_influence_1984,Cote_MacDonald_1991,Bonsall_Maradudin_1977,MacDonald_Murray_1985,dasilva2016fractional}. The recent observation of 
reentrant integer quantum Hall states~\cite{xu_signatures_2025,lu2024fractional,fci_nu_1by3} in both
rhombohedral graphene and TMD homobilayer systems suggests
that broken translational symmetry states are also ground states at some fractional band filling factors in these systems.

\begin{figure}
  \centering
  \includegraphics[width=\linewidth]{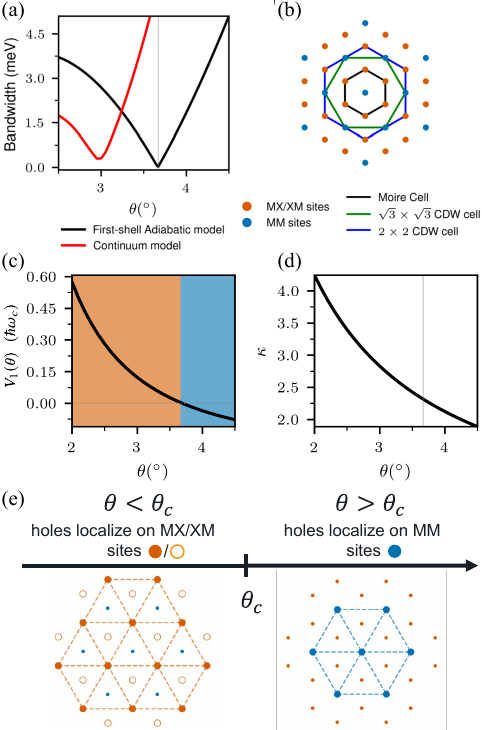}
  \caption{\label{fig:schematic}
Flat bands, real-space structure and effective moir\'e potential of tMoTe$_2$. 
(a)~Bandwidth (meV) versus twist angle of the Adiabatic model (first-shell approximation) and the Continuum model for tMoTe$_2$.
(b)~Schematic of the moiré unit cell illustrating MX, XM and MM sites, and  $\sqrt{3}\times \sqrt{3}$ and $2\times 2$ commensurate supercells.
(c)~Amplitude of the leading (first-shell) harmonic of the effective moir\'e potential $V_1(\theta)$ versus twist angle; $V_1$ changes sign at $\theta_c\simeq3.7^{\circ}$, switching the potential minima from MX (or XM, respectively) on the 
unit cell edges to MM at the unit-cell center. 
(d)~The Landau-level mixing parameter, $\kappa \equiv [e^2/\varepsilon\,\ell_B]/(\hbar\omega_c)$, versus twist angle for $\varepsilon = 10.0$. 
In (a,d), the vertical gray line marks $\theta_c\simeq3.7^\circ$.
(e)~Illustration of how the valence-band holes tend to localize on the MX or XM sites for $\theta<\theta_c$ and on the MM sites for $\theta>\theta_c$ in tMoTe$_2$. We use model parameters from Ref.~\cite{wang2024fractional}.}
\end{figure}

For tMoTe$_2$ moir\'es it is important to identify which charge-density-wave (CDW) patterns are favored as functions of twist angle and band filling, and how their energetics evolves across the magic-angle regime where FCI signatures are most prominent.  
In this Letter, we address these questions
by connecting the continuum model of tMoTe$_2$ to the well-understood physics of Landau levels in a periodic potential. We employ an adiabatic mapping~\cite{wu_topological_2019,shi_adiabatic_2024,PhysRevLett.132.096602,reddy2024non} that, when projected to a single LL, recasts the spin- and valley-polarized top valence band as an isolated Landau level subject to a triangular-lattice moiré potential.  
Our central finding is that the character of the CDW ground-state is largely controlled by the sign of the leading spatial harmonic of the effective potential, $V_1(\theta)$.   
The harmonic changes sign at a critical twist angle $\theta_c$ near the bandwidth minimum. 
The sign flip provides a simple selection rule: it exchanges the favored localization sites within the moir\'e unit cell from the MX/XM-stacked (metal-on-chalcogen) to the MM-stacked (metal-on-metal) regions [Fig.~\ref{fig:schematic}(a,b)]. 
Using large-scale, multi-LL self-consistent Hartree--Fock (HF) calculations, we 
show that this single rule organizes the landscape of
CDW ground states at commensurate fillings $\nu_h=1/4, 1/3, 1/2$, $2/3$ and $3/4$.  
At most filling factors, the CDW states can be 
identified as either electron or hole Wigner crystals pinned to the moiré cell high-symmetry site selected by the sign of $V_1$.

\textit{Continuum model and adiabatic mapping.---}
We model the valence-band moiré states of hole-doped tMoTe$_2$ with the two-layer continuum Hamiltonian~\cite{wu_topological_2019}
\begin{align}
H_{\mathrm{TMD}}
= -\frac{\hbar^2 k^2}{2m^*}\,\sigma_0
  + \Delta_0(\bm{r})\,\sigma_0
  + \bm{\Delta}(\bm{r})\!\cdot\!\bm{\sigma},
\end{align}
where the identity matrix $ \sigma_0$ and $ \bm{\sigma}$ are Pauli matrices in the layer pseudospin basis. 
We expand the moiré potentials $\Delta_0(\bm{r})$ and $\bm{\Delta}(\bm{r})$ in Fourier components using coefficients obtained from first-principles calculations for relaxed, corrugated tMoTe$_2$~\cite{wang2024fractional} and truncating at the first shell of moiré reciprocal vectors $ \{\bm{G}_j\}$. We 
assume below that interactions polarize the locked spin and valley 
degrees of freedom, as found experimentally over most of the regime of 
interest~\cite{cai2023signatures,ParkEtAl2025,chang2025evidence,li2025universal,xu_signatures_2025}, 
leaving a single active topological band with Chern number $ C=\pm 1$.

The adiabatic limit applies when
the pseudospin Zeeman energy \(|\bm{\Delta}(\bm{r})|\) is the dominant energy scale.
In this limit we apply a local unitary \(U(\bm{r})\) transformation that 
aligns the layer quantization axis with the pseudospin-field direction \(\hat{\bm{n}}(\bm{r})=\bm{\Delta}(\bm{r})/|\bm{\Delta}(\bm{r})|\), and project to the 
locally aligned layer pseudospin state.
The winding of $ \hat{\bm{n}}$ generates an emergent, spatially periodic vector potential \(\bm{\mathcal A} = \bm{\mathcal A}_0 + \delta \bm{\mathcal A}\) and a corresponding magnetic field \(\mathcal B(\bm{r})=\nabla\times\bm{\mathcal A}\). 
The latter has a uniform component $ \mathcal B_0$ corresponding to one flux quantum per moiré unit cell ($ \Phi_0\equiv hc/e$) and a periodic modulation \(\delta\mathcal B(\bm{r})\) contribution.
The effective single-component 
Hamiltonian~\cite{shi_adiabatic_2024,PhysRevLett.132.096602} 
\begin{align}
\label{eq:hadia}
H_{\mathrm{ad}}
&= -\frac{\hat{\bm{\Pi}}^2}{2m^*}
  +V,\\
    V &= -\frac{e}{2m^*c}\{\hat{\bm{\Pi}}, \delta\bm{\mathcal{A}}(\bm{r})\} - \frac{e^2}{2m^*c^2}(\delta\bm{\mathcal{A}}(\bm{r}))^2+ U_{\mathrm{ad}}(\bm{r}), \nonumber 
\end{align}
where $\hat{\bm{\Pi}}=-\mathrm{i}\hbar\nabla + e\bm{\mathcal{A}}_0/c$.  The scalar potential \(U_{\mathrm{ad}}(\bm{r})\) collects contributions from \(|\bm{\Delta}(\bm{r})|\), \(\Delta_0(\bm{r})\), and the quantum metric of the pseudospin texture. 
The uniform component $\mathcal B_0$ is conveniently treated by 
using the corresponding Landau-level representation for $H_{\mathrm{ad}}$. 
The field $ \mathcal B_0$ defines an effective magnetic length 
$\ell_B=\sqrt{\hbar c/(e\,\mathcal B_0)}$ and cyclotron energy \(\hbar\omega_c=e\,\mathcal B_0/(m^*c)\propto \theta^2\) . We quantify the 
interaction strength by the Landau-level mixing parameter \(\kappa \equiv [e^2/(\varepsilon\,\ell_B)]/(\hbar\omega_c)\propto 1/\theta\). 
When projected to the lowest Landau-level (LLL), $V$ acts as an effective scalar potential $V(\bm r)$, see Supplemental Material~\cite{SM} for details.
We denote the amplitude of its leading (first-shell) Fourier harmonic by
$V_1(\theta)$, see Fig.~\ref{fig:schematic}(c). 
The sign change of $V_1$ at the critical angle $\theta_c$ reverses the landscape of potential minima and therefore selects the preferred CDW pinning site.

\textit{Interactions and Hartree--Fock theory.---}
We study the influence of long-range Coulomb interactions,
\(H_{\mathrm{C}}=\frac{1}{2A}\sum_{\bm{q}} V_q :\hat n_{-\bm{q}}\hat n_{\bm{q}}:\),
using self-consistent Hartree–Fock (HF) theory, which provides
an accurate description of CDW states at fractional band fillings.
We employ the Landau gauge $\bm{\mathcal A}_0 = \mathcal B_0\,x\,\hat{\bm{y}}$, in which states within a Landau-level are labeled by guiding center $X$. The mean-field Hamiltonian is
\begin{align}
H_{\mathrm{HF}}
= \sum_{\alpha} \epsilon_\alpha\, c^\dagger_\alpha c_\alpha
+ \sum_{\alpha\beta}\!\left[ V + \Sigma^{\mathrm{HF}}[\Delta] \right]_{\alpha\beta} c^\dagger_{\alpha} c_{\beta},
\end{align}
where $ \alpha,\beta$ run over LL $n$ and guiding center $X$ indices.  The single-particle energies are \(\epsilon_{nX}\equiv \epsilon_n=-\hbar\omega_c\!(n+1/2)\),
$V$ is the single-particle moir\'e potential operator from Eq.~\ref{eq:hadia},
and $ \Sigma^{\mathrm{HF}}[\Delta]$ collects Hartree and Fock self-energy matrix elements, which are a linear function of the self-consistently determined guiding-center density matrix \(\Delta_{nn'}(\bm{G})\)~\cite{macdonald_influence_1984,Cote_MacDonald_1991}.
It is convenient to implement a particle–hole mapping so that the single-particle LL hole energies are positive (see SM~\cite{SM}).
We are interested in states with hole filling factor $\nu_h < 1$, but
both the single-particle operator $V$ and interactions mix Landau levels. 
We therefore include higher LLs, increasing the cutoff until energies and order parameters converge.  
At rational hole fillings $ \nu_h=p/q$, we search for moiré-commensurate CDW ground states [see Fig.~\ref{fig:schematic}(b)] by random seeding. 
We find ground states that either preserve $C_6$ rotational symmetry 
or break it down to $C_3$ or $C_2$ (stripe). 
For $ q=3$ and $ q=4$ we choose $ \sqrt{3}\!\times\!\sqrt{3}$ 
and $ 2\!\times\!2$ moiré supercells respectively \cite{note7}, 
since these periodicities allow CDW order to open a gap at the Fermi level.   

\textit{CDW properties.---}
When the moir\'e potential is ignored, our problem reduces to that of holes in 
Landau levels for which the CDW states are hole Wigner crystals (one hole per unit cell) when $\nu_h < 1/2$ and electron Wigner crystals for $\nu_h > 1/2$ \footnote{We work in a Hilbert space in which the microscopic degrees of freedom are holes.  By electron Wigner crystal we mean a state in which the CDW unit cell area contains one hole less than a full Landau level.  When Landau-level mixing is weak, these states are expected to be stable for $\nu_h > 1/2$ because electrons are then more readily localized than holes.}.
The HF ground-state energies $E_{\mathrm{tot}}(\theta)$ we obtain 
for $ \nu_h=1/3,\,2/3,\,1/4,\,1/2,$ and $ 3/4$ demonstrate that the moiré potential influences the CDW ground state. We first discuss $ \nu_h=1/3$ and $ 2/3$, for which we consider $ \sqrt{3}\!\times\!\sqrt{3}$ CDW supercells containing, respectively, one and two holes.

\begin{figure}[t]
  \centering
 \includegraphics[width=\linewidth]{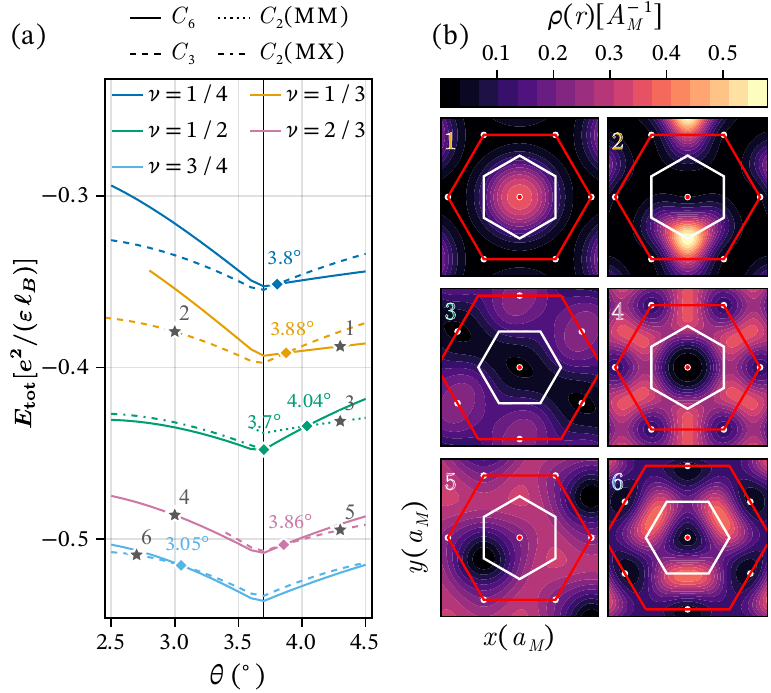}
  \caption{\label{fig:energy_competition}
Competing charge-ordered states over a range of hole fillings and twist angles. (a) Total energy per hole $E_{\textrm{tot}}$ versus twist angle $\theta$ for competing phases at fillings $\nu_h= 1/4,\, 1/3,\, 1/2,\, 2/3,\, 3/4$. The line style indicates the rotational symmetry of the corresponding state ($C_6$, $C_3$, $C_2$); first-order transition angles are marked by square markers. Numbered markers label configurations whose real-space densities are shown in (b).  (b) Real-space hole density $\rho_h(\bm{r})$ for the six labeled configurations in (a). White hexagons mark the moiré unit cell and red hexagons mark the CDW unit cell. 
Single-particle energies in (a) are relative to the twist-angle-dependent moiré band energy at $\nu_h=0$. 
}
\end{figure}

For ${\nu_h = 1/3}$, we find a transition near a critical angle $\theta_c$ between a $C_6$-symmetric CDW state and one that breaks the symmetry down to $C_3$ (see Fig.~\ref{fig:energy_competition}(a)).
When $ \theta<\theta_c$, the effective moiré potential favors localization on 
MX or XM sites (cf.~Fig.~\ref{fig:schematic}(c) and (e)).  For $\nu_h=1/3$ (one hole per CDW unit cell) the hole therefore occupies either the MX or XM sites, yielding a $ C_6$-broken hole Wigner crystal whose occupied sites form a triangular lattice. For $\theta>\theta_c$, the effective potential favors MM centers and produces a $C_6$-symmetric Wigner crystal. (see Fig.~\ref{fig:energy_competition}(b), panel 1.)
These CDW states can be viewed as Landau-level-like
Wigner crystals that are pinned to MX, XM, or MM sites, instead of having a gapless sliding mode, and are distorted differently in the three cases.

At $ \nu_h=2/3$ the trend reverses: two holes fill 
each $ \sqrt{3}\!\times\!\sqrt{3}$ CDW unit cell as illustrated in Fig.~\ref{fig:energy_competition}(a).  For $ \theta<\theta_c$, holes occupy both MX and XM sites and form a $ C_6$-symmetric state, leaving MM sites to be 
occupied by electrons (missing holes).  
(see Fig.~\ref{fig:energy_competition}(b), panel~4.) 
For $ \theta>\theta_c$, the moir\'e potential peaks at MX and 
XM sites and drives hole accumulation at MM, and either an MX or an XM site, lowering the symmetry from $ C_6\!\to\!C_3$, corresponding to electrons at XM or MX sites (see Fig.~\ref{fig:energy_competition}(b), panel~5.) \cite{note6}. 
Note that the energy difference between these crystal motifs is 
much smaller at $ \nu_h=2/3$ compared to at $ \nu_h=1/3$.
We attribute this difference to a reduction in the lattice-pinning energy in the 
electron crystal case that cannot take advantage of Landau-level 
mixing \footnote{At $\nu_h=1/3$ the crystal is a hole Wigner crystal in which the hole density $\rho_h(\bm{r})$ is maximized at lattice sites and increases with the Landau-level mixing parameter $\kappa$ (see Fig.~\ref{fig:schematic}(d)).  In contrast, at $ \nu_h=2/3$ the competing state is an electron CDW, corresponding to fully depleting $\rho_h(\bm{r})$ at lattice sites.
Because $\rho_h(\bm{r})\!\ge\!0$, the hole density at lattice sites is limited and 
LL mixing is less effective in further lowering energies.}.

\textit{Motif selection at other fillings.---}
For the $ 2\times2$ enlargement of the moir\'e unit cell, the same organizing principle --- the sign flip of $V_1(\theta)$ --- governs the commensurate cases summarized in Fig.~\ref{fig:energy_competition}. At $ \nu_h=1/4$ we find a sharp transition between two crystal motifs: for $ \theta<\theta_c$ the ground state is an MX/XM--centered triangular CDW, whereas for $ \theta>\theta_c$ it switches to an MM--centered triangular lattice.
At $ \nu_h=1/2$, the ground state for $ \theta>\theta_c$ is a stripe-ordered ($ C_2$-symmetric) CDW with holes localized near the MM sites (Fig.~\ref{fig:energy_competition}(b), panel 3), whereas for $ \theta<\theta_c$ it forms a honeycomb lattice with holes on the MX or XM sites of the enlarged unit cell.
At $ \nu_h=3/4$, there is a slight deviation in the pattern.  
The $ C_6$-symmetric and $ C_6$-broken solutions cross in energy
at an angle below $ \theta_c$, with the transition angle showing a strong dependence on interaction strength (parameterized by dielectric constant $ \varepsilon$, see SM). These results underscore the universal role of the effective moir\'e potential in selecting the ground-state geometry.

\begin{figure}
  \centering \includegraphics[width=0.95\linewidth]{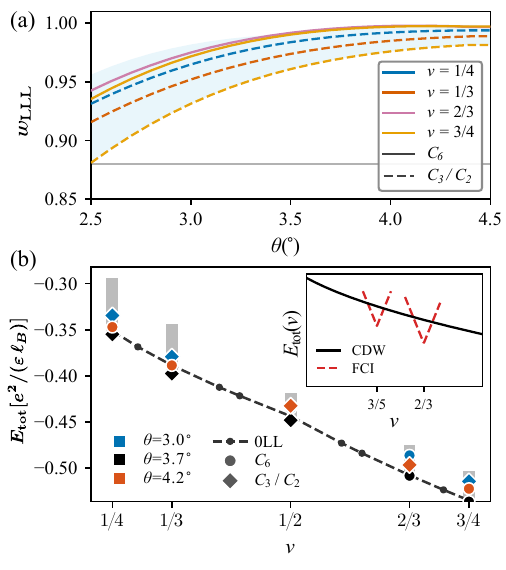}
\caption{\label{fig:LLLprojection}
  (a) Lowest-Landau-level weight $w_{\mathrm{LLL}}$ of the self-consistent CDW solutions versus twist angle $\theta$ (see SM~\cite{SM} for the definition of $w_{\mathrm{LLL}}$).
  The light-blue band shows the range across fillings $ \nu_h\in\{1/4,1/3,1/2,2/3,3/4\}$ 
  as a function of $\theta$. Representative curves are overlaid: color encodes filling $ \nu_h$, and line style encodes symmetry ($C_6$, $C_3$, $C_2$). The light-gray curve traces the minimum \(w_{\mathrm{LLL}}^{\mathrm{min}}(\theta)\) across all datasets. 
  (b) Total energy per hole versus filling at representative twist angles $\theta = 3.0^\circ$, $3.7^\circ$, and $4.2^\circ$; shaded bands cover the energy range over $\theta\in[2.5^\circ,4.5^\circ]$. Marker style indicates when $C_6$ symmetry is broken. The curve labeled ``$0$LL'' shows the interaction energy obtained by projection to the LLL (no Landau-level mixing) for comparison \cite{MacDonald_Murray_1985}.  Inset: schematic energetic competition between CDW and FCI states at $ \nu_h=2/3$ and $ 3/5$, analogous to Landau-level systems \cite{lam_liquid-solid_1984,levesque_crystallization_1984}.
  }
\end{figure}

\textit{Filling-factor dependence of the ground state energy.---} 
In Fig.~\ref{fig:LLLprojection}(b) we plot the ground-state energy versus 
$\nu_h$ for representative twist angles; the shaded band indicates the spread over $ \theta\in[2.5^\circ,4.5^\circ]$. The magnitudes and trends are consistent with Hartree–Fock studies of interactions in Landau levels~\cite{macdonald_influence_1984,Yoshioka_Lee_1983,yoshioka1979charge}, reinforcing that the physics in tMoTe$_2$ closely parallels that of Landau levels over a broad range of twists \cite{kousa2025theory}.  
Figure~\ref{fig:LLLprojection}(a) reports the lowest-Landau-level (LLL) weight $ w_{\mathrm{LLL}}$, defined as the LLL occupation fraction \(w_{\mathrm{LLL}}\equiv (1/\nu_h N_\phi)\,\mathrm{Tr}(P_{\mathrm{LLL}}\rho)\), see SM~\cite{SM} for details. Across all twist angles and fillings studied we find $ w_{\mathrm{LLL}}\ge 0.88$, indicating that most of the charge resides in the LLL while Landau-level mixing lowers the energy by enabling enhanced spatial localization. Moreover, $ w_{\mathrm{LLL}}$ decreases with decreasing $ \theta$, consistent with the $ \theta$-dependence of the LL-mixing parameter $ \kappa$ in Fig.~\ref{fig:schematic}(d). This accounts for the more localized charge-density patterns at smaller $ \theta$ [Fig.~\ref{fig:energy_competition}(b), panels 4 and 6].

\textit{Miniband Chern Numbers.---} 
For each self-consistent solution we compute the total Chern~\cite{macdonald1984quantized,vanderbilt2018berry}
summed over occupied Hartree--Fock minibands:
\begin{align}
    C = \frac{1}{2\pi} \int_{\mathrm{mBZ}} \!\!\!\!\!\! d^2 k 
    \sum_{\alpha} n_{\alpha\bm{k}}\, 2\,\mathrm{Im}\!
    \left\langle \partial_{k_x} u_{\alpha\bm{k}} \,\middle|\, \partial_{k_y} u_{\alpha\bm{k}} \right\rangle .
\end{align}
Here $n_{\alpha\bm{k}}$ is the occupation of band $\alpha$ at pseudomomentum $\bm{k}$, and the magnetic cell-periodic states are
$|u_{\alpha\bm{k}}\rangle = e^{-i\bm{k}\cdot\hat{\bm{r}}}|\psi_{\alpha\bm{k}}\rangle$, see SM~\cite{SM}. 
States at $\nu_h\le 1/2$ are topologically trivial, whereas those at $\nu_h>1/2$ carry $|C|=1$, except 
for the $C_6$-broken state labeled 6 in Fig.~\ref{fig:energy_competition}(a,b), which is topologically trivial ($C=0$) because strong Landau-level mixing localizes the holes.

\textit{Discussion.---}
We have shown that the charge-density-wave states in twisted MoTe$_2$
can be viewed as pinned hole Wigner crystals at $\nu_h<1/2$ and as 
pinned electron Wigner crystals at $\nu_h>1/2$, with crystal distortions that 
are dependent on the favored pinning site within the moir\'e period.
In some $\nu_h$ and twist angle regimes, the electron density is polarized toward 
XM or MX sites, implying that the CDW state is spontaneously layer polarized,
suggesting that its response to displacement fields could be hysteretic.
In this case we would also expect that the filling factor range 
over which CDW states are stable relative to FCI states should be 
enhanced by weak displacement fields.  The electron Wigner-crystal 
states have mean-field bands with nonzero total Chern number, and therefore 
provide a natural explanation for the reentrant integer quantum Hall (RIQH) states 
observed experimentally \cite{xu_signatures_2025} in this system,
as we explain more fully below.  
Although the Hartree--Fock mean-field method we employ provides 
a good description of CDW states in twisted MoTe$_2$ flat bands, it cannot
describe the fractional Chern insulator states (FCI) with which they compete.
Based on experiment~\cite{chang2025evidence,xu_signatures_2025,chen2025fractional,park2023observation,ParkEtAl2025}, it appears that FCI states are stabilized at Jain-sequence fractional filling factors (e.g., $\nu_h=3/5$ and $2/3$)~\cite{jain1989composite}. 
We interpret the RIQH states seen experimentally as (weak-moiré pinned) electron Wigner-crystal CDW states, and suggest that crystalline phases can be stabilized over a broader range of fractional fillings than in conventional Landau-level 2DEGs, where CDWs are typically observed over continuous intervals only near $\nu=0$ and $\nu=1$ (Wigner crystals) or in narrow intermediate-filling windows (often discussed near $\nu\simeq 1/4$).
It is in fact expected \cite{kousa2025theory,MacDonald_Murray_1985} that the effective moir\'e potential should stabilize CDW states relative to FCI states because their energy is lowered at first order in the potential, compared to second order in the FCI case.  

The CDW states we find can be viewed as partial Hall crystals~\cite{tevsanovic1989hall}, i.e., translation-symmetry-broken states that (in the electron-crystal regime) can retain an integer quantized Hall response.
In the effective Landau-level description used throughout this work, the Landau-level degeneracy per area is set by the total effective flux density.
With one flux quantum per moir\'e unit cell from the emergent field and an external magnetic field $B$ (defined with sign relative to the emergent field), this degeneracy density is
$(A_0^{-1}+B/\Phi_0)$, where $A_0$ is the moir\'e unit-cell area and $\Phi_0$ is the flux quantum.
For an electron Wigner crystal (one electron per broken-symmetry unit cell of area $A^{\rm EC}$), the hole density $p^*$ is then the Landau-level degeneracy density minus the electron density, yielding
\begin{equation}
p^* \;=\; \frac{1}{A_0}+\frac{B}{\Phi_0}-\frac{1}{A^{\rm EC}}.
\end{equation}
The coefficient of $B$ in this relation is sometimes referred to as the St\v{r}eda slope~\cite{Streda_1982}.
As noted elsewhere~\cite{xu_signatures_2025,huang2025apparent}, the magnetic-field dependence of $p^*$ extracted experimentally does not follow a fixed-commensuration form like the equation above, and instead exhibits an \emph{apparent} St\v{r}eda slope intermediate between those of the competing FCI states at bracketing filling factors.
This behavior is expected if the system responds to changes in magnetic field by changing the crystal unit-cell area $A^{\rm EC}$ (or by switching between nearby commensurations), and it requires that Wigner-crystal energies are less sensitive to $\nu_h$ than the competing FCI states, as indicated schematically in the inset of Fig.~\ref{fig:LLLprojection}(b).
As $B$ is varied, a sequence of reentrant integer quantum Hall Wigner-crystal states with differing $A^{\rm EC}$ can then be stabilized over the interval of $\nu_h$ between the $\nu_h=2/3$ and $\nu_h=3/5$ FCI St\v{r}eda lines, where CDW states compete most effectively.

Indeed, our calculations show that the CDW energies closely approximate those of electron and hole Wigner crystals in Landau levels, whose energies depend smoothly on $\nu_h$ and admit arbitrary unit-cell areas.
Similar remarks apply to the apparent slopes of plateau boundaries for stripe and bubble states in Landau levels, which are controlled by competition with nearby incompressible FQH states and therefore need not directly measure the Hall conductivity.
Transitions between Wigner crystal and FCI states, which are apparently abundant in the experimental phase diagram, must be first order because of the topological distinctions between the correlations and quasiparticle charged excitation spectra of the two systems.  

In experiments, disorder is essential to establish robust quantum Hall plateaus. To leading (linear) order in the disorder potential, we 
expect that disorder will also stabilize CDW states relative to FCI/FQH states~\cite{kousa2025theory,girvin1986magneto,endo2010collapse}, implying that increasing disorder enlarges the parameter region occupied by CDWs while shrinking the window for FCIs~\cite{li2010observation}. 
A quantitative comparison of ground-state energies, including disorder averaging, to map the CDW--FCI competition more precisely, is beyond the scope of this Letter, but is a natural next step for future studies. 

\textit{Acknowledgments.--}
We thank Tingxin Li, Jingtian Shi, Nicolas Morales-Durán, Bo Zou and Ze-xun Lin for helpful discussions.  This work was supported by a Simons Foundation Targeted Grant under Award No. 896630.

\bibliography{refs}


\clearpage
\onecolumngrid

\setcounter{secnumdepth}{2}
\setcounter{equation}{0} 
\setcounter{figure}{0} 
\setcounter{table}{0} 
\setcounter{page}{1}
\makeatletter
\renewcommand{\theequation}{S\arabic{equation}}
\renewcommand{\thefigure}{S\arabic{figure}}
\renewcommand{\thetable}{S\arabic{table}}
\makeatother

\setlength{\parindent}{0pt} 
\setlength{\parskip}{6pt plus 2pt minus 1pt} 

\section*{Supplementary Material}
Throughout the Supplementary Material we switch from Gaussian to SI units. We follow notation and definitions from \cite{Cote_MacDonald_1991,macdonald_influence_1984}. 
\section{Derivation of the Adiabatic model}
\subsection{Continuum model}
We follow Refs.~\cite{shi_adiabatic_2024,PhysRevLett.132.096602,li2025variational} closely for this section. Homobilayer TMDs can be described by a two-layer effective model written in the layer basis \cite{wu_topological_2019}:
\begin{eqnarray}
    H_{TMD}=\begin{pmatrix}
        -\frac{\hbar^2}{2\,m^*}({\hat{\bm{k}}}-{\bm k}_b)^2+\Delta_b({\bm r})&\Delta_T({\bm r})\\
        \Delta_T^{\dagger}({\bm r})&-\frac{\hbar^2}{2\,m^*}({\hat{\bm{k}}}-{\bm k}_t)^2+\Delta_t({\bm r})
    \end{pmatrix}.
\end{eqnarray}
Here, $\hat{\bm {k}}  = -i \nabla$ is the wavenumber operator and $\bm k_{b/t}$ are the bottom and top momentum shifts corresponding to the bottom and top TMD layer $K$ (or $K'$) points. $\Delta_{b/t}(\bm r )$ are the intra-layer potentials and $\Delta_T (\bm r)$ is the inter-layer tunneling term.
We do a layer-dependent gauge transformation:
\begin{eqnarray}
    H_{TMD} \rightarrow  \begin{pmatrix}
        e^{-i \bm{k}_b \cdot \bm{r}} & 0\\
        0 & e^{-i \bm{k}_t \cdot \bm{r}}
    \end{pmatrix} H_{TMD} \begin{pmatrix}
        e^{i \bm{k}_b \cdot \bm{r}} & 0\\
        0 & e^{i \bm{k}_t \cdot \bm{r}}
    \end{pmatrix}.
\end{eqnarray}
This makes the resulting Hamiltonian quasi-periodic in the moir{\'e} unit cell (periodic up to a unitary transformation, note that the spectrum remains periodic). Under this transformation:
\begin{align}
    \Delta_T(\bm{r}) \rightarrow e^{i \bm q _1 \cdot \bm r}\Delta_T (\bm{r}),\quad \bm q_ 1 = \bm k_t - \bm k_b.
\end{align}
We then get:
\begin{eqnarray}
        H_{\text{TMD}}&=\begin{pmatrix}
        -\frac{\hbar^2}{2\,m^*}{ \hat{\bm {k}}}^2+\frac{\Delta_b-\Delta_t}{2}&\text{Re}\,\Delta_T({\bm r})+i\,\text{Im}\, \Delta_T({\bm r})\\
        \text{Re}\,\Delta_T({\bm r})-i\,\text{Im}\, \Delta_T({\bm r})& -\frac{\hbar^2}{2\,m^*}{\hat{\bm{k}}}^2-\frac{\Delta_b-\Delta_t}{2}
    \end{pmatrix}+\begin{pmatrix}
    \frac{\Delta_b(\bm r)+\Delta_t(\bm r)}{2}&0\\[0.2 in]
            0&\frac{\Delta_b(\bm r)+\Delta_t(\bm r)}{2}
        \end{pmatrix} \nonumber \\
        &=-\frac{\hbar^2\,{\hat{\bm {k}}}^2}{2m^*}\,\sigma_0+{\bm \Delta}({\bm r})\cdot {\bm \sigma}+\Delta_0({\bm r})\,\sigma_0,\quad \bm \Delta = \left( \text{Re} \Delta_T^{\dagger}, \text{Im} \Delta_T^{\dagger}, (\Delta_b - \Delta_t)/2\right). \label{eq:TMDcont}
\end{eqnarray}
Where now:
\begin{eqnarray}
    \Delta_0(\bm r ) =  \frac{\Delta_b(\bm r)+\Delta_t(\bm r)}{2}.
\end{eqnarray}
We focus on the first shell approximation for the intralayer potentials and interlayer tunneling:
\begin{align}
    \Delta_{b/t}(\bm r) = 2 V \sum_{j = 1,3,5} \cos(\bm G _j \cdot \bm{r} \pm \psi),\quad \Delta_T = \omega \sum _{i=1}^3 e^{i \bm q_i \cdot \bm r}.
\end{align}
Where, $\bm{G}_{j} = C_6^j \bm{G}_{1}$ for $j = 1\dots 6$ and  $\bm{G}_{1} = k_\theta (1/2,\sqrt{3}/2)$. We define $\bm q_{i+1} = C_3^i \bm q_1$, $i=0,1,2$ and use twist angle parameters obtained from large-scale DFT calculations that were done for the twist angle $3.89^\circ$.
\subsection{Adiabatic model}
 We perform a \textit{local} unitary transformation at each $\bm r$ such that it diagonalizes the pseudospin texture term in Eq.~(\ref{eq:TMDcont}):
\begin{align}
    U^\dagger (\bm r ) \left(\bm \Delta (\bm r ) \cdot \bm \sigma\right )U (\bm r ) = |\bm \Delta (\bm r )| \sigma _z.
\end{align}
If the local eigenvectors of the operator are given by $\chi_\pm(\bm r )$:
\begin{align}
    \left(\bm \Delta (\bm r ) \cdot \bm \sigma\right ) \chi_\pm(\bm r ) = \pm |\bm \Delta (\bm r )| \chi_\pm(\bm r ),
\end{align}
Then, the  unitary operator that diagonalizes the operator is:
\begin{align}
U(\bm r)=
\begin{pmatrix}
\big| & \big| \\
\bm \chi_+(\bm r) & \bm \chi_-(\bm r) \\
\big| & \big|
\end{pmatrix}.
\end{align}
Performing this transformation on the Hamiltonian Eq.~(\ref{eq:TMDcont}) gives us:
\begin{align}
    H' =  U^\dagger (\bm r ) H U (\bm r ) &= U^\dagger (\bm r ) \left(\frac{-\hbar^2\,{\hat{\bm {k}}}^2}{2m^*}\right)\,U (\bm r )\sigma_0+U^\dagger (\bm r ) \left(\bm \Delta (\bm r ) \cdot \bm \sigma\right )U (\bm r ) +\Delta_0({\bm r})\,\sigma_0 \\
    &= \frac{-\hbar^2}{2m^*}U^\dagger (\bm r ) \hat{\bm {k}}^2 \,U (\bm r ) + |\bm \Delta (\bm r )|\sigma_z +\Delta_0(\bm r )\sigma_0 \\
    &= \frac{-\hbar^2}{2m^*} \left( \hat{\bm {k}}  -i U^\dagger (\bm r )\nabla  U (\bm r )\right)^2 + |\bm \Delta (\bm r )|\sigma_z +\Delta_0(\bm r )\sigma_0\\
    &= \frac{-\hbar^2}{2m^*} \left( \hat{\bm {k}} \sigma_0 +   \dfrac{e}{\hbar } \bm A (\bm r )\right)^2 + |\bm \Delta (\bm r )|\sigma_z +\Delta_0(\bm r )\sigma_0.
\end{align}
Where we define the U(2) gauge field $\bm{A}(\bm{r})$ \cite{onishi2026emergent} with components:
\begin{align}
     A_i (\bm r ) =  \frac{-i\hbar }{e} \;  U^\dagger (\bm r ) \partial_i U (\bm r ) =  \frac{-i\hbar }{e} \begin{pmatrix}
        \bm \chi_+^\dagger \partial_i \bm \chi_+   &\bm \chi_+^\dagger \partial_i \bm \chi_-\\
        \bm \chi_-^\dagger \partial_i \bm \chi_+ & \bm \chi_-^\dagger \partial_i \bm \chi_-
    \end{pmatrix}\equiv\begin{pmatrix}
        {A}_{\uparrow \uparrow,i}& {A}_{\uparrow \downarrow,i}\\
         {A}_{\uparrow\downarrow,i }^*& {A}_{\downarrow \downarrow,i}
    \end{pmatrix}.
\end{align}
Expanding the $\bm{A}(\bm{r})$ terms we get \cite{PhysRevLett.132.096602}:
\begin{align}
    H' = \begin{pmatrix}
        -\dfrac{1}{2m^*}\left (\hbar \hat{\bm {k}}  + e \bm{A}_{\uparrow \uparrow} \right)^2 - \dfrac{e^2}{2m^*}  |\bm{A}_{\uparrow \downarrow}|^2+ |\bm{\Delta}(\bm r )| + \Delta_0(\bm r) & -\frac{\hbar e}{2m^*}\{\hat{\bm{k}},\bm A_{\uparrow \downarrow}\} -\frac{e^2}{2m^*} \bm{A}_{\uparrow \downarrow}(\bm{A}_{\uparrow \uparrow} + \bm{A}_{\downarrow \downarrow})\\
        -\frac{\hbar e}{2m^*}\{\hat{\bm{k}},\bm A_{\downarrow \uparrow }\} -\frac{e^2}{2m^*} \bm{A}_{ \downarrow \uparrow}(\bm{A}_{\uparrow \uparrow} + \bm{A}_{\downarrow \downarrow}) & -\dfrac{1}{2m^*}\left (\hbar \hat{\bm {k}}  + e \bm{A}_{\downarrow \downarrow } \right)^2 - \dfrac{e^2}{2m^*} |\bm{A}_{\uparrow \downarrow}|^2 - |\bm{\Delta}(\bm r )| + \Delta_0(\bm r)
    \end{pmatrix} .
\end{align}
We now project the Hamiltonian to the low-energy sector denoted by $\uparrow$. This is valid under the assumption that the kinetic energy fluctuations are small $\hbar^2 /2m^*A_M \ll |\bm{\Delta}(\bm r)|$. We then obtain the \textit{Adiabatic Hamiltonian}:
\begin{align}
    H_{\mathrm{ad}} &\equiv P_\uparrow H' P_\uparrow = -\frac{\hat{\bm{\Pi}}^2}{2m^*}
  +V,\\
    V &= -\frac{e}{2m^*}\{\hat{\bm{\Pi}}, \delta\bm{\mathcal{A}}(\bm{r})\} - \frac{e^2}{2m^*}(\delta\bm{\mathcal{A}}(\bm{r}))^2+ U_{\mathrm{ad}}(\bm{r}) \label{eq:def_V},\\
    U_{\mathrm{ad}}(\bm{r}) &= - \dfrac{e^2}{2m^*}  |A_{\uparrow \downarrow}|^2+ |\bm{\Delta}(\bm r )| + \Delta_0(\bm r), \\
    \bm{A}_{\uparrow \uparrow } &= \bm {\mathcal{A}}_0 + \delta \bm{\mathcal{A}},\quad \nabla \times \bm{\mathcal{A}}_0 = \mathcal{B}_0 \hat{\bm{z}},  \quad \nabla \times \delta \bm{\mathcal{A}} = \delta \mathcal{B}(\bm{r}) \hat{\bm{z}}.
\end{align}
where $\hat{\bm{\Pi}}=-\mathrm{i}\hbar\nabla + e\bm{\mathcal{A}}_0$. 
Within the topological regime of tMoTe$_2$ we have:
\begin{align}
     |\mathcal{B}_0| A_M/\Phi_0 = 1, \quad \delta \mathcal{B}(\bm r + \bm R ) = \delta \mathcal{B}(\bm r ),\quad \int_{A_M} d^2 r \; \delta \mathcal{B}(\bm r)=0.
\end{align}
The scalar potential \(U_{\mathrm{ad}}(\bm{r})\) collects contributions from \(|\bm{\Delta}(\bm{r})|\), \(\Delta_0(\bm{r})\), and the quantum metric of the pseudospin texture. The pseudomagnetic field is decomposed into a spatially constant part $\mathcal{B}_0$ with one flux quantum per moir{\'e} unit cell and into $\delta \mathcal{B}$, a spatially periodic (pseudo)magnetic field with zero spatial average within the moir{\'e} unit cell. Note that Eq. (\ref{eq:hadia}) of the main text is written in Gaussian units. 

\subsection{Effective potential}
In the main text we define $V_1$ as the amplitude of the leading (first-shell) Fourier harmonic of the operator $V$ in Eq.~(\ref{eq:def_V}) after projection to the lowest Landau level (LLL)~\cite{PhysRevLett.132.096602}. We detail the projection here. The operator $V$ contains the mechanical momentum operator $\hat{\bm\Pi}$ and is therefore not a purely local function of $\bm r$ (unlike $U_\mathrm{ad}(\bm r)$ and $\delta\bm{\mathcal A}(\bm r)$). Nevertheless, the LLL-projected operator is fully determined by its matrix elements within the LLL, and these can be reproduced by a local potential with the same projected matrix elements.

We rewrite Eq.~(\ref{eq:def_V}) in terms of $\Pi_\pm=\Pi_x\pm i\Pi_y$ and $\mathcal A_\pm=\mathcal A_x\pm i\mathcal A_y$:
\begin{align}
V
= -\frac{e}{2m^*}\Big(
\Pi_+\delta\mathcal A_-+\Pi_-\delta\mathcal A_+
\Big)
-\frac{e^2}{2m^*}\,\delta\mathcal A_+\,\delta\mathcal A_-
+U_\mathrm{ad},
\label{eq:V_pm_form}
\end{align}
where we have used Coulomb gauge, $\nabla\!\cdot\!\delta\bm{\mathcal A}(\bm r)=0$, to simplify the symmetrized minimal-coupling term.
We expand the local fields in plane waves,
\begin{align}
U_\mathrm{ad}(\bm r)=\sum_{\bm G}U_\mathrm{ad}(\bm G)e^{i\bm G\cdot\bm r},
\qquad
\delta\mathcal A_\pm(\bm r)=\sum_{\bm G}\alpha_\pm(\bm G)e^{i\bm G\cdot\bm r},
\label{eq:fourier_defs}
\end{align}
and express $\Pi_\pm$ in terms of Landau-level ladder operators for electrons in a magnetic field $+\hat{\bm z}$:
\begin{align}
\Pi_+ = i\frac{\sqrt{2}\hbar}{\ell_B}\,a^\dagger,
\qquad
\Pi_- = -i\frac{\sqrt{2}\hbar}{\ell_B}\,a.
\label{eq:Pi_ladder}
\end{align}
(Our $+\hat{\bm z}$ convention differs from Sec.~\ref{sec:LL_operators}, which uses $-\hat{\bm z}$; see Refs.~\cite{pfannkuche_1992,PhysRevLett.132.096602} for consistent conventions.)

To perform the LLL projection we use the standard Landau-gauge identity for plane-wave matrix elements between adjacent Landau levels,
\begin{align}
\langle 1X'|e^{i\bm G\cdot\bm r}|0X\rangle
= i\frac{G_-\ell_B}{\sqrt{2}}\,
\langle 0X'|e^{i\bm G\cdot\bm r}|0X\rangle,
\label{eq:pw_identities}
\end{align}
with $G_\pm=G_x\pm iG_y$. Combining Eqs.~(\ref{eq:V_pm_form})--(\ref{eq:pw_identities}), the projected matrix elements can be written as
\begin{align}
\langle 0X'|V|0X\rangle
= \sum_{\bm G} V_{\bm G}\,
\langle 0X'|e^{i\bm G\cdot\bm r}|0X\rangle,
\label{eq:Vproj_def}
\end{align}
which are identical to those generated by a local potential
\begin{align}
V(\bm r)=\sum_{\bm G}V_{\bm G}\,e^{i\bm G\cdot\bm r}.
\label{eq:Veff_local}
\end{align}
The corresponding Fourier coefficients are
\begin{align}
V_{\bm G}
=
-\frac{\hbar e}{2m^*}\,G_-\,\alpha_+(\bm G)
-\frac{e^2}{2m^*}\sum_{\bm G'}\alpha_+(\bm G')\alpha_-(\bm G-\bm G') + U_\mathrm{ad}(\bm G).
\label{eq:Veff_coeff}
\end{align}

Equations~(\ref{eq:Vproj_def})--(\ref{eq:Veff_coeff}) provide a convenient definition of the leading moir\'e harmonic within the LLL: $V_1(\theta)$ is the amplitude of the first-shell component of $V(\bm r)$, and inherits its twist-angle dependence through $U_\mathrm{ad}$ and $\delta\bm{\mathcal A}$. This quantity is plotted in Fig.~\ref{fig:schematic}(c) of the main text. The sign change of $V_1(\theta)$ corresponds to an inversion of the effective potential landscape. In the HF calculations of the main text, we retain only the first-shell Fourier harmonics for the potentials; the model is not restricted to the LLL.

\section{Landau-level operators}\label{sec:LL_operators}

The following relations are useful and consistent with electrons in a magnetic field $\bm{B}  = (0,0,-B)$ (equivalently, holes in $\bm{B}  = (0,0,B)$) for $B>0$ \cite{zou2025valley}:
\begin{align}
    \Pi_x &= p_x + e A_x , \quad \Pi_y = p_y + e A_y,\quad [\Pi_i, \Pi_j ] = +i \hbar^2 \ell_B^{-2} \epsilon_{ij},\quad \ell_B = \sqrt{\hbar/eB}, \\
    a &= i\frac{\ell_B}{\hbar\sqrt{2}}\left(\Pi_x + i \Pi_y \right)  ,\quad a^\dagger = -i\frac{\ell_B}{\hbar\sqrt{2}}\left(\Pi_x - i \Pi_y \right),\quad [a,a^\dagger] = 1,[a^\dagger a,a^\dagger] = a^\dagger,\\
      \Pi_x &= i\frac{\hbar}{\ell_B \sqrt{2}} (a - a^\dagger ),\quad \Pi_y =\frac{\hbar}{\ell_B \sqrt{2}} (a^{\dagger} + a ),
    \\
    H_0 &= \hbar\omega_c \left(a^\dagger a + \frac{1}{2} \right ),\quad \omega_c = eB/m^*,\\
    \bm{R} &= \bm{r} - \frac{\ell_B^2}{\hbar} \hat{z} \times \bm{\Pi} ,\quad [R_i, R_j] = -i \ell_B ^2 \epsilon_{ij}, \quad [R_i, \Pi_j] = 0,\quad [R_i,H_0] = 0,\\
    b &= \frac{1}{\sqrt{2}\ell_B} (R_x + i R_y),\quad b^\dagger = \frac{1}{\sqrt{2}\ell_B} (R_x - i R_y),\quad [b,b^\dagger] = 1,\quad [b,a] =[b^\dagger,a] =0,\\
     \bm{K} &= \bm{\Pi} + \hbar \ell_B^{-2} (\hat{z} \times \bm{r}) = + \hbar \ell_B^{-2} (\hat{z} \times \bm{R}),\quad [K_i,K_j] = -i\epsilon_{ij}\hbar^2 \ell_B ^{-2}, \\
     T_{\bm{a}} &= e^{i\bm{a}\cdot \bm{K}/\hbar},\quad   T_{\bm{a} }T_{\bm{a}'} =   T_{\bm{a}'}  T_{\bm{a}} e^{ 2\pi i n_\phi },\quad n_\phi = \frac{ B |\bm{a} \times \bm{a}'|}{ \Phi_0 } 
\end{align}

\subsection{Plane-wave matrix elements}
The plane-wave matrix elements can be written as \cite{macdonald_influence_1984,Cote_MacDonald_1991,mishra_2025_17807688,zou2025valley}
\begin{align}
    \bra{n'X'}e^{i\bm{q}\cdot\bm{r}}\ket{nX} = \bra{0X'}e^{i\bm{q}\cdot\bm{R}}\ket{0X} \bra{n0}e^{-i \bm{q}\cdot\frac{\ell_B^2}{\hbar} \hat{z} \times \bm{\Pi}}\ket{n'0},
\end{align}
and for the guiding-center and Landau-level matrix elements, one finds 
\begin{align}
\bra{X'}e^{i \bm{q}\cdot\bm{R}}\ket{X}=e^{i q_x (X+X')/2} \delta_{X',X+q_y \ell_B^2}, 
&&
\bra{n'}e^{i \bm{q}\cdot\frac{\ell_B^2}{\hbar} \hat{z} \times \bm{\Pi}}\ket{n}=(F_{\bm{q}})_{n'n}
=\left(\frac{n!}{n'!}\right)^{\frac{1}{2}} \left(\frac{(-q_y+iq_x)\ell_B}{\sqrt{2}} \right)^{n'-n} e^{-\frac{q^2\ell_B^2}{4}} L_n^{n'-n}\!\left(\frac{q^2\ell_B^2}{2}\right),
\end{align}
for $n\leq n'$, where $L_n^{\alpha}(x)$ is the generalized Laguerre polynomial and $(F_{\bm{q}})_{nn'} =  (F_{-\bm{q}})_{n'n}^*$.

\section{Commensurability of CDW and moir\'e patterns in a magnetic field}
Given a filling $\nu$ of the moir\'e band (triangular lattice), we enumerate the CDWs compatible with a given flux per CDW unit cell.

Let $n_\phi^M$ and $n_\phi^{\mathrm{CDW}}$ denote the magnetic flux per moir\'e and CDW unit cell, respectively. The topological non-trivial winding of the texture leads to $n_\phi^M=1$. We focus on commensurate periodicities that break translation symmetry while preserving $C_6$ rotational symmetry, implying a larger real-space CDW unit-cell area and $n_\phi^{\mathrm{CDW}}>n_\phi^M$. Assuming rational flux per CDW unit cell,
\begin{align}
    n_\phi^{\mathrm{CDW}} = \frac{p}{q} = \frac{B A_{\mathrm{CDW}}}{\Phi_0} 
    = \frac{B}{n_{\mathrm{CDW}} \Phi_0}\, n_{\mathrm{CDW}}A_{\mathrm{CDW}}
    = \frac{B A_{\mathrm{tot}}}{N_{\mathrm{tot}} \Phi_0}\, n_{\mathrm{CDW}}A_{\mathrm{CDW}},
\end{align}
where we have written the number of holes per CDW unit cell as $ n_{\mathrm{CDW}}A_{\mathrm{CDW}}$. Since $\nu = N_{\mathrm{tot}}/N_{\phi,\mathrm{CDW}}$,
\begin{align}
    n_\phi ^{\mathrm{CDW}} = \frac{n_{\mathrm{CDW}}A_{\mathrm{CDW}}}{\nu}.
\end{align}
Because $A_{\mathrm{CDW}}>A_M$, $n_\phi ^{\mathrm{CDW}} >1$. For the Brillouin zones, $BZ_{\mathrm{CDW}} < BZ_{M}$.

\subsection{CDW reciprocal lattice vectors}
Let $ \bm{g}^{\mathrm{CDW}}_1 $ and $\bm{g}^{\mathrm{CDW}}_2=C_6\,\bm{g}^{\mathrm{CDW}}_1$ be CDW reciprocal basis vectors, and $ \bm{g}^{M}_1, \bm{g}^{M}_2$ the moir\'e ones. If commensurate,
\begin{align}
    \bm{g}^{M}_1 = s_1 \bm{g}^{\mathrm{CDW}}_1 + r_1 \bm{g}^{\mathrm{CDW}}_2,
    \qquad
    \bm{g}^{M}_2 = s_2 \bm{g}^{\mathrm{CDW}}_1 + r_2 \bm{g}^{\mathrm{CDW}}_2,
\end{align}
with $\bm{g}^{\mathrm{CDW}}_2 = C_6\bm{g}^{\mathrm{CDW}}_1$ so that $s_2,r_2$ can be expressed in terms of $s_1, r_1$.
Since the two reciprocal lattices are commensurate, the moir\'e vectors differ from those of the CDW by a rotation $R_\theta$ and a multiplicative constant $\alpha$:
\begin{align}
    \bm{g}^{M}_1 = \alpha R_{\theta} \bm{g}^{\mathrm{CDW}}_1 ,\quad \bm{g}^{M}_2 = C_6 \bm{g}^{M}_1.
\end{align}
Hence
\begin{align}
    (\alpha R_\theta - s_1)\bm{g}^{\mathrm{CDW}}_1 = r_1 C_6 \bm{g}^{\mathrm{CDW}}_1 
    \quad\Rightarrow\quad
    \alpha^2 = r_1 ^2 + s_1 ^2 + r_1 s_1.
\end{align}
Note that $\alpha^2 = |g_M|^2/|g_{\mathrm{CDW}}|^2 = n_\phi^{\mathrm{CDW}}/n_\phi^{M} = n_{\mathrm{CDW}}A_{\mathrm{CDW}}/\nu$.
Thus,
\begin{align}
    r_1 ^2 + s_1 ^2 + r_1 s_1 = n_\phi^{\mathrm{CDW}} \left(=\frac{n_{\mathrm{CDW}}A_{\mathrm{CDW}}}  {\nu}\right).
\end{align}
Integer solutions $(r_1,s_1)\in\mathbb Z^2$ exist iff $n_\phi^{\mathrm{CDW}}\in\mathbb Z$. Hence commensurate CDWs require integer flux per CDW unit cell.

\section{Particle--hole transformation}
In practice, we apply a particle--hole transformation to convert holes in a fully filled band to electrons in an empty band. For any system, transforming position-space operators as $\psi_h^\dagger(\bm{r}) = \psi_e(\bm{r})$ yields
\begin{align}
    H = \int d^2 \bm{x}  \, \psi_h(\bm{x})h_0(\bm{x}) \psi_h^\dagger(\bm{x}) = \operatorname{tr} h_0 + \int d^2 \bm{x} \, \psi_h^\dagger(\bm{x})(-h_0(\bm{x}))^* \psi_h(\bm{x}).
\end{align}
For the adiabatic model the particle--hole transformed Hamiltonian is
\begin{align}
    H_{\text{ad}} \rightarrow H_{\text{ad}}  &= \frac{(\hat{\bm{p}} + e\bm{\mathcal{A}})^2}{2m^*} - V^*(\bm{r})\\
    &=\frac{(\hat{\bm{p}} + e\bm{\mathcal{A}}_0)^2}{2m^*}  +\frac{1}{2m^*}\{(\hat{\bm{p}} + e\bm{\mathcal{A}}_0), \delta\bm{\mathcal{A}}(\bm{r})\} + \frac{(\delta\bm{\mathcal{A}}(\bm{r}))^2}{2m^*} - U_{\text{ad}}(\bm{r}) .
\end{align}
The transformation maps holes in $+\hat{z}$ field to electrons in $-\hat{z}$ field. We perform the self-consistent HF procedure in Landau gauge $\mathcal{A}_0 = (0, -B  x)$ for $\bm{B} = -B \hat z$ with $B>0$. 

\section{Hartree--Fock matrix elements}
The Hamiltonian in the LL basis is
\begin{align}
H
&=  \sum_{nX,\, n'X'}  c_{n X}^\dagger [H_0]_{nX,n'X'} c_{n' X'}
\;+\; \frac{1}{2}\sum_{\{n_i X_i \}}
\big\langle n_1 X_1, n_2 X_2 \big| V_c \big| n_1' X_1', n_2' X_2' \big\rangle\;
c_{n_1 X_1}^\dagger c_{n_2 X_2}^\dagger c_{n_2' X_2'} c_{n_1' X_1'}.
\end{align}
Throughout we assume long-range Coulomb interactions:
\begin{align}
    V(\bm{r} ) = \frac{e^2}{4\pi \varepsilon_0 \varepsilon r},\qquad V_{\bm{q}} = \frac{ e^2}{2 \varepsilon_0 \varepsilon q}.
\end{align}

Normal ordering with respect to the ground state reads
\[
: c_{n_1 X_1}^\dagger c_{n_2 X_2} :\,
= c_{n_1 X_1}^\dagger c_{n_2 X_2}
- \big\langle c_{n_1 X_1}^\dagger c_{n_2 X_2} \big\rangle .
\]
The Hartree--Fock Hamiltonian is
\begin{align}
H\simeq H_{HF} 
&\equiv \sum_{nX,\, n'X'}  c_{n X}^\dagger [H_0]_{nX,n'X'} c_{n' X'}
\\
&+ \sum_{\{n_i X_i \}}
\big\langle n_1 X_1, n_2 X_2 \big| V_c \big| n_1' X_1', n_2' X_2' \big\rangle
\left(
c_{n_1 X_1}^\dagger c_{n_1' X_1'} \,\big\langle c_{n_2 X_2}^\dagger c_{n_2' X_2'} \big\rangle
-\; \big\langle c_{n_1 X_1}^\dagger c_{n_2' X_2'} \big\rangle\, c_{n_2 X_2}^\dagger c_{n_1' X_1'} \right)\\
&- \frac12\sum_{\{n_i X_i \}} \big\langle n_1 X_1, n_2 X_2 \big| V_c \big| n_1' X_1', n_2' X_2' \big\rangle
\Big(
\langle c_{n_1 X_1}^\dagger c_{n_1' X_1'} \rangle \,\big\langle c_{n_2 X_2}^\dagger c_{n_2' X_2'} \big\rangle - \langle c_{n_1 X_1}^\dagger c_{n_2' X_2'} \big\rangle\,\langle c_{n_2 X_2}^\dagger c_{n_1' X_1'} \rangle 
\Big) .
\end{align}
The interaction part can be written as
\begin{align}
H_{\text{int}}^{\text{HF}}
&= \sum_{\{n_i X_i \}}
\Big(
\big\langle n_1 X_1, n_2 X_2 \big| V_c \big| n_1' X_1', n_2' X_2' \big\rangle
\,\big\langle c_{n_2 X_2}^\dagger c_{n_2' X_2'} \big\rangle
\;-\;
\big\langle n_1 X_1, n_2 X_2 \big| V_c \big| n_2' X_2', n_1' X_1' \big\rangle
\,\big\langle c_{n_2 X_2}^\dagger c_{n_2' X_2'} \big\rangle
\Big)\;
c_{n_1 X_1}^\dagger c_{n_1' X_1'} \\
&= \sum_{\{n_i X_i \}} \left(\Sigma^H(n_1 X_1,n_1' X_1')-\Sigma^X(n_1 X_1,n_1' X_1')\right)c_{n_1 X_1}^\dagger c_{n_1' X_1'}.
\end{align}
We write the Hartree and exchange self-energies as
\begin{align}
    \Sigma^H(n_1 X_1,n_1' X_1') &= \frac{e^2}{4\pi \varepsilon_0  \varepsilon \ell_B}\sum_{\bm{G}}\left(e^{iG_x(X_1 + X_1')/2}\delta_{X_1,X_1' + G_y \ell_B^2}\right) \sum_{n_2 n_2'} U_H(n_1,n_1';n_2,n_2'|\bm{G})\Delta_{n_2,n_2'}(\bm{G}),\\
    \Sigma^X(n_1 X_1,n_1' X_1') &= \frac{e^2}{4\pi \varepsilon_0 \varepsilon \ell_B} \sum_{\bm{G}}\left(e^{iG_x(X_1 + X_1')/2}\delta_{X_1,X_1' + G_y \ell_B^2}\right) \sum_{n_2 n_2'} U_X(n_1,n_1';n_2,n_2'|\bm{G})\Delta_{n_2,n_2'}(\bm{G}),
\end{align}
with
\begin{align}
    U_H(n_1,n_1';n_2,n_2'|\bm{G})= \frac1{G \ell_B} F_{n_1 n_1'}(\bm{G})F_{n_2 n_2'}(-\bm{G}),
    && 
    U_X(n_1,n_1';n_2,n_2'|\bm{G})= \int d^2 (\bm{q} \ell_B) \frac{e^{i (\bm{q} \times \bm{G})\ell_B^2}}{q \ell_B}\, F_{n_1 n_2'}(\bm{q})F_{n_2 n_1'}(-\bm{q}).
\end{align}
We evaluate the exchange term analytically and implement the resulting closed-form expression in numerics \cite{macdonald_influence_1984} (Note: numerical implementations can also be found \cite{mishra_2025_17807688}).
Total, single-particle, and interaction energies are
\begin{align}
    E_{\mathrm{sp}} \equiv \mathrm{tr}(H_0 \rho),\qquad
    E_{\mathrm{tot}} = \mathrm{tr}(H_{0}\rho) + \frac{1}{2}\, \mathrm{tr}(\Sigma^{\rm HF}\rho),\qquad
    E_{\mathrm{int}} = E_{\mathrm{tot}} - E_{\mathrm{sp}}.
\end{align}

\subsection{Density matrix and relations} \label{supmatsec:density_matrix_relations}
Define the operator $\hat \rho_{nn'}(X,X') \equiv c_{n'X'}^\dagger c_{nX}$ 
whose ground-state expectation value is the density matrix,
\begin{align}
   \rho_{nn'}(X,X') \equiv  \langle c_{n'X'}^\dagger c_{nX}\rangle =\sum_\alpha n_\alpha \langle nX|\alpha \rangle \langle \alpha | n' X'\rangle.
\end{align}
The real-space density operator $\hat\rho(\bm{r})
= \Psi^\dagger(\bm{r}) \Psi(\bm{r})$ and its Fourier transform $\hat{\rho}_{\bm{q}}
\equiv \int_{A_\text{sys}}d^2 r \, e^{-i\bm{q}\cdot \bm{r}} \, \hat \rho(\bm{r})$ are
\begin{align}
    \hat\rho(\bm{r})
= \sum_{nX,n'X'} \phi_{n'X'}^*(\bm{r}) \, \phi_{nX}(\bm{r}) \,
\hat\rho_{nn'}(X,X'), 
&&
\hat{\rho}_{\bm{q}}
= \sum_{nX,n'X'}
\langle n'X' | e^{-i\bm{q}\cdot \bm{r}} | nX \rangle \,
\hat \rho_{nn'}(X,X')
= \sum_{n'n} F_{n'n}(-\bm{q}) \,
\hat \Delta_{n'n}(\bm{q}),
\end{align}
where we used $\langle n'X' | e^{i\bm{q}\cdot \bm{r}} | nX \rangle
= \delta_{X',\,X + \ell_B^2 q_y}\,
e^{i q_x (X+X')/2}\, F_{n'n}(\bm{q})$ (see above) and the guiding-center density matrices
\begin{align}
   \hat \Delta_{n'n}(\bm{q})
=\frac{1}{N_\phi} \sum_X e^{-i q_x X} \, e^{i q_x \ell_B^2 q_y/2} \,
\hat \rho_{nn'}(X,X - \ell_B^2 q_y)
&&
\left[\text{so that: }
  \hat \rho_{nn'}(X, X')=  \sum_{\bm{q}} e^{i q_x (X + X')/2}\delta_{X',X-q_y \ell_B^2}\hat \Delta_{n' n}(\bm{q}) \right] .
\end{align}

In the CDW ground state characterized by reciprocal lattice vectors $\bm{G}$, we have
\begin{align}
    \langle\hat \rho_{\bm{q}}\rangle  = \delta_{\bm{q}, \bm{G}}\rho_{\bm{G}},
\qquad
\langle \hat \Delta_{n' n}(\bm{q}  )  \rangle = \delta_{\bm{q}, \bm{G}}\,\Delta_{n' n}(\bm{G}) .
\end{align} 

\subsection{Sum rule for the lowest Landau level}
We note that 
\begin{align}
    \Delta_{n'n}(\bm{G})
= \frac{1}{N_\phi}\,
   \operatorname{tr}_{X}\!\left(e^{-i \bm{G} \cdot \bm{R}}\,\rho_{nn'}\right) 
\qquad
\Longrightarrow
\qquad
\sum_{\bm{G}} \Delta_{n'n}(\bm{G})\,\Delta^{*}_{m'm}(\bm{G})
= \frac{1}{N_\phi}\,
   \operatorname{tr}_X\!\left(\rho_{nn'}\,\rho_{mm'}\right).
\end{align}

When truncating to the LLL,
\begin{align}
    \sum_{\bm{G}} |\Delta_{00}(\bm{G})|^2 = \frac{1}{N_\phi}\,
   \operatorname{tr}_X\!\left(\rho_{00}\,\rho_{00}\right) = \frac{1}{N_\phi}\,
   \operatorname{tr}_X\!\left(\rho_{00}\right) = N/N_\phi = \nu,
\end{align}
the usual LLL sum rule \cite{macdonald_influence_1984,yoshioka1979charge,Yoshioka_Lee_1983}. With Landau-level mixing, $\sum_{\bm{G}} |\Delta_{00}(\bm{G})|^2<\nu$.

\section{Lowest-Landau-level weight}
To quantify how much a self-consistent solution resides in the LLL, define the lowest Landau-level weight $w_{\mathrm{LLL}}$:
\begin{align}
w_{\mathrm{LLL}}
\equiv \frac{1}{\nu}\,\frac{1}{N_\phi}\operatorname{Tr}\!\left(P_{\mathrm{LLL}}\rho\right),
\qquad 0\le w_{\mathrm{LLL}}\le 1,
\end{align}
where $ P_{\mathrm{LLL}}$ projects onto the LLL, $ \rho$ is the self-consistent HF single-particle density matrix, $ N_\phi$ is the LL degeneracy, and \(\nu \equiv (1/N_\phi)\operatorname{Tr}\rho\) is the filling factor.
Because \(\sum_n \Delta_{nn}(\bm{0})=\nu\) and
\(\Delta_{00}(\bm{0})= (1/N_\phi)\,\operatorname{tr}_X(e^{-i\bm{0}\cdot\bm{R}}\rho_{00})\),
\begin{align}
w_{\mathrm{LLL}}
= \frac{1}{\nu}\,\Delta_{00}(\bm{0}).
\end{align}

\section{Basis, matrix elements, and numerical details}

For clarity, the main text expresses operators in the guiding-center basis of Landau gauge, $ \{|n,X\rangle\}$. For bulk calculations and improved speed/accuracy, our simulations use a Bloch-like $k$-space basis following Ref.~\cite{macdonald1984quantized}:
\begin{align} \label{eq:quasibloch_rep}
    |k_x k_y; n i\rangle
    = \frac{1}{\sqrt{S}}
       \sum_{l\in\mathbb Z} e^{ i k_x X_{i l}(k_y)}\, |n, X_{i l}(k_y)\rangle,
    \qquad
    X_{i l}(k_y)
    = \ell_B^2\!\left[k_y + Q_2\big(i + l\,j\big)\right],
\end{align}
where $ i\in\{0,1,\dots, j-1\}$ labels the $ j$ magnetic sublattices within the moir\'e magnetic unit cell, and $ \ell_B$ is the magnetic length. The normalization constant is
\begin{equation}
    S \;=\; \frac{L\, Q_0\, p}{2\pi\, j\, q},
\end{equation}
with flux density $ n_\phi=p/q$ (coprime $ p,q\in\mathbb Z^+$) and
\begin{equation}
    j \;=\; \frac{p\,t}{\gcd(t,q)}.
\end{equation}
The reciprocal lattice vectors are
\begin{equation}
    \bm{b}_1=(Q_0,0),\qquad \bm{b}_2=(Q_1,Q_2),\qquad \frac{Q_1}{Q_0}=\frac{s}{t},
\end{equation}
with coprime integers $ s,t$. For a triangular lattice, $s=1$ and $t=2$. We take the magnetic Brillouin-zone ranges to be
\begin{equation}
    k_y\in(0,Q_2],\qquad
    k_x\in\Big(0,\; \frac{Q_0\,p}{j\,q}\Big].
\end{equation}

\subsection{Plane-wave matrix elements}
Writing a reciprocal vector as $ \bm{G}=a\,\bm{b}_1+b\,\bm{b}_2$ with $ a,b\in\mathbb Z$, the plane-wave matrix elements in the $ |k_x k_y; n i\rangle$ basis are
\begin{align}
    \langle k_x' k_y'; n' i' | e^{i\bm{G}\cdot \bm{r}} | k_x k_y; n i \rangle
    &= \delta_{k_y,k_y'}\,\delta_{k_x,k_x'}\, F_{n'n}(\bm{G})\,
       e^{ -i Q_2\, b\, k_x \ell_B^{2} }\,
       e^{ i \ell_B^{2} G_x (k_y + G_y/2) }\,
       e^{ i G_x \ell_B^{2} Q_2 i }\,
       \delta_{\,i',\, (i+b)\bmod j}.
\end{align}
Equivalently, in dimensionless variables \(\tilde{k}_x = k_x /(Q_0 n_\phi/j)\), $ \tilde{k}_y = k_y/Q_2$, and $ \tilde G_x = G_x/Q_0$,
\begin{align}
    \langle k_x' k_y'; n' i' | e^{i\bm{G}\cdot \bm{r}} | k_x k_y; n i \rangle
    &= \delta_{k_y,k_y'}\,\delta_{k_x,k_x'}\,\delta_{\,i',\, (i+b)\bmod j}\,
       F_{n'n}(\bm{G})\,
       e^{ -2\pi i\, b\, \tilde k_x / j }\,
       e^{ 2\pi i\, \tilde G_x (\tilde k_y + b/2)/n_\phi }\,
       e^{ 2\pi i\, \tilde G_x i / n_\phi } .
\end{align}
Here \(F_{n'n}(\bm{G})\) is the usual LL form factor.
The number of magnetic unit cells along $ \hat y$ is
\begin{equation}
    S=\frac{L}{Q_2\, j\, \ell_B^2} = \frac{L\, Q_0\, p}{2\pi\, j\, q}.
\end{equation}

\subsection{Numerical parameters and convergence}
Energies are reported in units of \(e^2/(4\pi\varepsilon_0\varepsilon\,\ell_B)\). We systematically increased the LL cutoff until the HF ground-state energy converged to four decimal places; a cutoff of 10 LLs was sufficient for convergence for both the single-particle spectrum of the adiabatic model and the HF ground-state energy. For Brillouin-zone sampling, we used an $ 11\times 11$ $ k$-grid to obtain converged density matrices and doubled the grid in the final iteration when reporting ground-state energies. Further increases in either LL cutoff or $ k$-grid size did not change the energy beyond the fourth decimal place, indicating negligible finite-size effects at our working parameters.

Figure~\ref{fig:energy_competition} compiles results for several symmetry-broken states. Not all ans\"atze admit self-consistent solutions across the full twist-angle range; where a branch fails to converge, it is omitted.

\subsection{Chern number}
To compute the Chern number for a Fermi energy within a target gap we define the magnetic cell-periodic functions corresponding to the magnetic Bloch states $|\bm{k};n,i\rangle$ with $i\in\{0,1,\dots ,j-1\}$ and $j = pt/\gcd(q,t)$:
\begin{align}
|u_{\bm{k}};n,i\rangle = e^{-i \bm{k} \cdot \hat{\bm{r}}}|\bm{k};n,i\rangle 
\qquad
\Longrightarrow
\qquad
   \langle u_{\bm{k}};n,i|u_{\bm{k}'};n',i'\rangle =  \langle \bm{k} ;n,i|e^{i (\bm{k} - \bm{k}') \cdot \bm{r}}|\bm{k}' ;n',i'\rangle.
\end{align}
With \cref{eq:quasibloch_rep} we then find
\begin{align}
     \langle \bm{k} ;n,i|e^{i (\bm{k} - \bm{k}') \cdot \bm{r}}|\bm{k}' ;n',i'\rangle  
     &= \delta_{ i ,i'} F_{nn'}(\bm{k}- \bm{k}')  e^{-i\pi (\tilde k_x + \tilde k_x') (\tilde k_y - \tilde k_y')/j},\quad  k_x = \tilde k_x Q_0 p/(qj),\quad k_y = \tilde k_y Q_2.
\end{align}
After the HF step, we have $H|\bm{k},\alpha\rangle = \epsilon_{\bm{k} \alpha} |\bm{k},\alpha\rangle$ with
\begin{align}
|\bm{k},\alpha \rangle = \sum_{n, i}C^\alpha _{n,i} |\bm{k}; n,i\rangle ,
&&
    |u_{\bm{k} \alpha}\rangle = \sum_{n, i}C^\alpha _{n,i} e^{-i \bm{k}\cdot \hat{\bm{r}}}|\bm{k}; n,i\rangle, 
&&
\Longrightarrow
&&
    \langle u_{\bm{k}\alpha}| u_{\bm{k}'\alpha'} \rangle  = (\bm{C}^\alpha _{\bm{k}})^\dagger \mathrm{PW}(\bm{k}, \bm{k}')\bm{C}^{\alpha'} _{\bm{k}'},
\end{align}
where $\mathrm{PW}(\bm{k}, \bm{k}')$ is the matrix of plane-wave overlaps above. We then evaluate the non-Abelian Berry curvature on a discrete $k$-mesh to obtain $C$~\cite{vanderbilt2018berry,fukui2005chern}.
\section{Dielectric-constant-dependent phase transition at $\nu_h = 3/4$}

\begin{figure}
    \centering
    \includegraphics[width=0.75\linewidth]{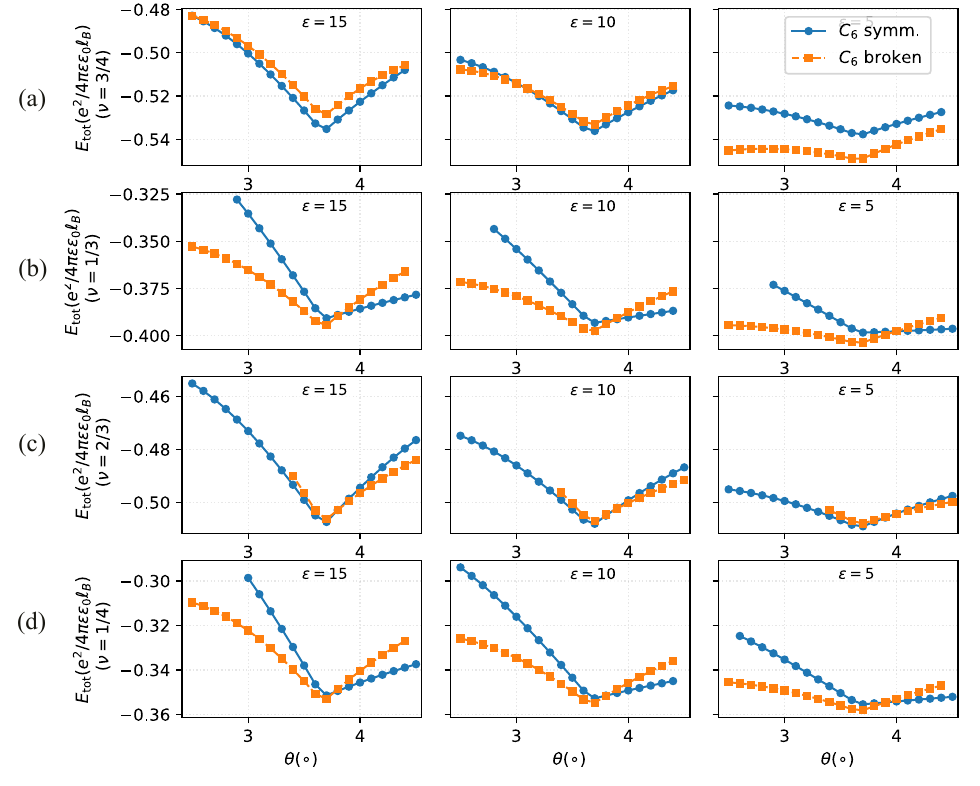}
    \caption{Total ground-state energy for dielectric constant values $\varepsilon=15,10,5$ (first, second, and third columns from the left) in units of $e^2/(4\pi \varepsilon \varepsilon_0 \ell_B)$ for fillings (a) $3/4$, (b) $1/3$, (c) $2/3$, and (d) $1/4$.}
    \label{fig:dielectric_ct_plots}
\end{figure}

Here we discuss the dielectric-constant dependence of the ground-state energy at $\nu_h = 3/4$. We find a pronounced $\varepsilon$-dependence of the energetically preferred lattice pinning configuration. For stronger Coulomb interactions (smaller $\varepsilon$), the $C_6$-symmetry-broken state in which the \textit{electron} is pinned to the MX or XM site has lower energy for all twist angles in $[2.5^\circ,4.5^\circ]$, whereas for weaker interactions (larger $\varepsilon$) the transition shifts towards smaller twist angles as shown in Fig.~\ref{fig:dielectric_ct_plots}(a). Such a strong dependence was not seen for $\nu_h = 1/3$, $2/3$, and $1/4$ [Fig.~\ref{fig:dielectric_ct_plots}(b), (c), and (d), respectively], where the twist angle at which the crossover occurs shifts by only $\sim 0.5^\circ$.

\end{document}